\documentclass[fleqn,twoside]{article}
\usepackage{espcrc2}
\usepackage{graphicx}
\def\ne{{n_e}}
\def\dne{{\delta n_e}}
\def\DM{{\rm DM}}
\def\RM{{\rm RM}}
\def\SM{{\rm SM}}
\def\EM{{\rm EM}}
\def\bpar{{B_{\parallel}}}

\newcommand{\arcsec}{\ensuremath{{}^{\prime\prime}}}

\newcommand{\hii}{H\,{\sc ii}}
\newcommand{\hi}{H\,{\sc i}}
\def\cnsq{{C_n^2}}

\def\ths{{\theta_\mathrm{src}}}
\def\thd{{\theta_d}}
\def\thij{{\theta_{ij}}}
\def\thiso{{\theta_{\mathrm{iso}}}}

\def\rlc{{r_{\rm LC}}}

\newcommand{\lesssim}{\mathbin{\lower 3pt\hbox
   {$\rlap{\raise 5pt\hbox{$\char'074$}}\mathchar"7218$}}} 
\newcommand{\gtrsim}{\mathbin{\lower 3pt\hbox
   {$\rlap{\raise 5pt\hbox{$\char'076$}}\mathchar"7218$}}} 
\title{The Microarcsecond Sky and Cosmic Turbulence}
\author{T.~Joseph~W.~Lazio\address[NRL]{Naval Research Laboratory,
		Washington, DC, USA; Joseph.Lazio@nrl.navy.mil}
		\thanks{Basic research in radio astronomy at the NRL
		is supported by the Office of Naval Research.}
	J.~M.~Cordes\address[CU]{Cornell University and NAIC, Ithaca, NY, USA;
		cordes@astro.cornell.edu}
		\thanks{This work was supported by NSF grants to
		Cornell University, AST~9819931, AST~0138263, and
		AST~0206036 and also by the National Astronomy \&
		Ionosphere Center, which operates the Arecibo
		Observatory under a cooperative agreement with the \hbox{NSF}.}
	A.~G.~de~Bruyn\address[ASTRON]{ASTRON \& Kapteyn Astronomical
		Institute, The Netherlands; ger@astron.nl}
	J.-P.~Macquart\address[RUG]{Kapteyn Astronomical Institute,
		University of Groningen, The Netherlands;
		jpm@astro.rug.nl}}

\begin{document}

\begin{abstract}
Radio waves are imprinted with propagation effects from ionized media
through which they pass.  Owing to electron density fluctuations,
compact sources (pulsars, masers, and compact extragalactic sources)
can display a wide variety of scattering effects.  These scattering
effects, particularly interstellar scintillation, can be exploited to
provide \emph{superresolution}, with achievable angular resolutions
($\lesssim 1\,\mu$arc sec) far in excess of what can be obtained by
very long baseline interferometry on terrestrial baselines.
Scattering effects also provide a powerful \emph{sub-AU} probe of the
microphysics of the interstellar medium, potentially to spatial scales
smaller than 100~km, as well as a tracer of the Galactic distribution
of energy input into the interstellar medium through a variety of
integrated measures.  Coupled with future $\gamma$-ray observations,
SKA observations also may provide a means of detecting fainter compact
$\gamma$-ray sources.  Though it is not yet clear that propagation
effects due to the intergalactic medium are significant, the SKA will
either detect or place stringent constraints on intergalactic
scattering.
\end{abstract}

\maketitle

\section{Introduction}\label{sec:turb.intro}

All radio observations of Galactic and extragalactic objects are
conducted while viewing these objects through the Galaxy's
interstellar medium (ISM).  Dispersion of pulsar signals and optical
observations, particularly of the H$\alpha$ emission line
\cite{hrtmjp03}, indicate that the interstellar medium contains a
diffuse ionized component in addition to classical H\,\textsc{ii}
regions.  This interstellar plasma occupies a significant fraction
($\sim 0.2$) of the volume of the \hbox{ISM}, and the energy required
to keep it ionized is considerable, roughly 15--20\% of the luminosity
of all of the O stars in the Galaxy.  The interstellar plasma is also
known as the Warm Ionized Medium (WIM) or the Diffuse Ionized Gas
(DIG).  Significantly for the \hbox{SKA}, the interstellar plasma will
produce observable effects on every line of sight over at least a
fraction of the entire proposed operating frequency range.

The SKA can allow unprecedented use of interstellar scattering and
interstellar scintillation (ISS) for study of sources and intervening
media:

\begin{itemize}
\item Resolving source structure on angular scales $\lesssim
10^{-6}$~arcsec will probe the inner jet regions in active galactic nuclei and individual source
regions in pulsar magnetospheres.  Astrophysically, such observations may
be crucial to an understanding of acceleration of particles in these
and other kinds of sources.  The SKA's sensitivity over a wide range
of frequencies is needed to exploit the resolving power of \hbox{ISS}.

\item Probing microturbulence in the ionized ISM through
sampling of scattering phenomena along large numbers ($\gtrsim 10^4$)
lines of sight.  The microturblence plays important roles in the
energetics of the ISM and in the propagation of cosmic rays in the
Galaxy's magnetic field.

\item Probing the intergalactic medium (IGM) through detection of angular
broadening in excess of that expected from foreground Galactic
gas. The SKA allows angular broadening measurement of large samples of
active galactic nuclei (e.g., $10^6$ lines of sight) that will map out
contributions from individual intervening galaxies, galaxy clusters,
Ly-$\alpha$ clouds, and from a pervasive but clumpy ionized IGM that
comprises most of the baryonic matter in the Universe.
\end{itemize}

Radio-wave scattering from density fluctuations in the interstellar
plasma offers both a means of characterizing the ISM on sub-AU scales
as well as a powerful probe of source structure.  In the remainder of
this section, we describe observable effects and astrophysical
measures to quantify the interstellar plasma.  Subsequent sections
discuss interstellar scintillation and superresolution of source
structure (\S\ref{sec:turb.structure}), the microphysics of the
interstellar plasma (\S\ref{sec:turb.physics}), interstellar
scattering as a means to measure distances
(\S\ref{sec:turb.distribute}), the possibility of intergalactic
scattering (\S\ref{sec:turb.igm}), the potential impact of
scintillations on SKA observations (\S\ref{sec:turb.problems}), and
key observations to be carried out with the SKA
(\S\ref{sec:turb.summary}).

\subsection{Observable Effects}\label{sec:turb.effects}

\begin{figure}
\includegraphics[width=\columnwidth]{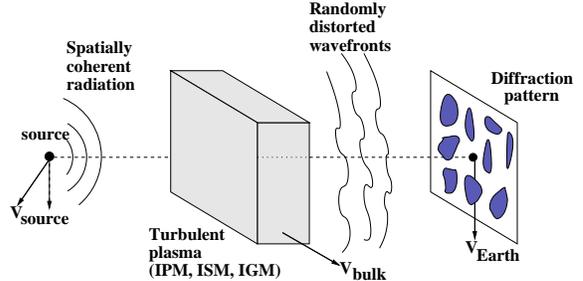}
\vspace{-1.25cm}
\caption{Geometry of radio-wave scattering in the ISM that is
responsible for a variety of seeing and scintillation effects.}
\label{fig:turb.geometry}
\end{figure}

The phase of a radio wave of wavelength~$\lambda$ after having
propagated through a plasma is 
\begin{equation}
\phi = \int dz\,\lambda\,r_en_e(z),
\label{eqn:turb.phase}
\end{equation}
where $r_e$ is the classical electron radius and~$n_e$ is the plasma
density.  Soon after the discovery of pulsars, it became clear that
their signals were displaying intensity scintillations
(e.g., \cite{r70}), akin to the scintillation of star light at
optical wavelengths viewed through the Earth's atmosphere.  Intensity
scintillations result from constructive and destructive interference
and occur if the phase of the wave has been perturbed during its
propagation with structure on small transverse scales.  
As eqn.~(\ref{eqn:turb.phase}) shows, phase
perturbations imply that the propagating wave has been scattered by
interstellar electron density fluctuations.  Figure~\ref{fig:turb.geometry}
illustrates the relevant geometry.

Since the detection of pulsar intensity scintillations, a rich variety
of additional interstellar scattering observables has been recognized.
These include \cite{r90,cl91}:
\begin{description}
\item[Intensity Scintillation:] The original indication of interstellar
density fluctuations, intensity scintillations have been detected not
only from pulsars but potentially also from active galactic nuclei
(AGN) and from masers.

\item[Angular broadening (``seeing''):] Compact radio sources,
both Galactic and extragalactic, show angular broadening by 
amounts ranging from $\lesssim 1$~mas to~1~arcsec at~1~GHz, depending
on the distance and direction.  The angular smearing scales
approximately as $\nu^{-2}$.

\item[Pulse broadening:] Multipath propagation causes a multiplicity
of arrival times, usually seen as an exponential-like ``tail'' to pulses
from pulsars.

\item[Angular wandering:] Focussing and defocussing of rays by
AU-scale structures can cause apparent shifts in source positions.

\item[Pulse time of arrival (TOA) fluctuations:]  Changes in geometry caused
by proper motions of a pulsar and intervening material induce
variations in the amount of plasma along the line of sight.  Also,
variable scattering causes variable arrival times.

\item[Spectral Broadening:]  Broadening of narrow spectral lines
from a combination of scattering and time-variable geometry.  This
effect is very small in the ISM ($\lesssim 1$ Hz) and has not been
found because there are no known sources that produce spectral lines
narrow enough for this effect to be significant.  It has been measured
from spacecraft viewed through the interplanetary medium \cite{hc83}, and, if
extraterrestrial transmitters are ever found, it may be important for
them.
\end{description}

\subsection{Integrated Measures}\label{sec:turb.measures}

As we discuss in more detail below,
radio measurements provide four integrals involving the electron density
and magnetic field.   Differential arrival times from pulsars provide the 
\emph{dispersion measure}, 
\begin{eqnarray}
\DM &=& \int_0^D ds\, \ne(s)
        \quad ({\rm  pc\, cm^{-3}}), 
\end{eqnarray}
(for $D$ in pc and $n_e$ in cm$^{-3}$).  A magnetized plasma produces
Faraday rotation, which is described by the \emph{rotation measure},
\begin{eqnarray}
\RM &=& \frac{e^3}{2\pi m_e^2c^4} \int_0^D ds\, \ne\bpar,\nonumber\\
    &=& 0.81\,\mathrm{rad\,m^{-2}}\, \int_0^D ds\, \ne\bpar,
\label{eqn:turb.rm}
\end{eqnarray}
the integral of the magnetic field component parallel to the line-of-sight
$\bpar$ (in $\mu$G) and which is sensitive to the correlation of
electron density and magnetic field. 

The \textit{emission measure} can be measured from 
recombination line and free-free absorption or emission observations
in the radio and observations of H$\alpha$ in the optical
\begin{eqnarray}
\EM = \int_0^D ds\, \ne^2
        \quad ({\rm  pc\, cm^{-6}})
\end{eqnarray}
(for $D$ in pc and $n_e$ in cm$^{-3}$).

As we discuss below, the density fluctuations responsible for
interstellar scattering can be described in terms of a power spectrum.
The \textit{scattering measure} is the integral of $\cnsq$, the
coefficient of the wavenumber spectrum for electron-density
fluctuations, $\dne$,
\begin{eqnarray} 
\SM = \int_0^D ds\, \cnsq(s) \quad (\mathrm{kpc\,m}^{-20/3}).
\label{eqn:turb.sm}
\end{eqnarray}
Cordes et al.~\cite{cwfsr91} give relations between \hbox{DM},
\hbox{SM}, and \hbox{EM}.

Finally, diffuse $\gamma$-ray emission is given by
\begin{equation}
I_\gamma = \int ds\,n_n(s)q_\gamma(s),
\label{eqn:turb.g-rayI}
\end{equation}
where $n_n$ is the nucleon density which is dominated by the hydrogen
number density~$n_H$, i.e., \hii, \hi, and~H${}_2$; $q_\gamma$ is the
gamma-ray emissivity per nucleon, which is proportional to the
cosmic-ray density; and $ds$ is a path-length element along the line
of sight.  Although not yet exploited, combining future radio and
$\gamma$-ray observations may allow fluctuations in $q_\gamma$,
resulting from magnetic field fluctuations, to be traced.

From large sets of measurements of these measures, but, in particular
\DM, \SM, and \RM, the spatial distribution and volume filling factor
of the electron density and its fluctuations, e.g., as parameterized
by a wavenumber spectrum for quantities such as $\dne$ and
$\delta\bpar$, can be probed as can the large-scale structure and
fluctuations in the magnetic field.

\section{Interstellar Scintillation (ISS) and Superresolution}\label{sec:turb.structure}

``Stars twinkle but planets do not'' is an exploitation of how
\emph{superresolution} is obtained via propagation through a turbulent
medium.  Planets do not twinkle, even though their reflected light
travels through the same turbulent atmosphere as does starlight,
because their angular diameters are so large as to quench the
scintillations.  Comparison of a twinkling star and non-twinkling
planet obtains constaints on their angular diameters that are roughly
an order of magntiude better than the nominal resolution of the human
eye.  ``Pulsars twinkle but AGNs do not'' is the interstellar
equivalent, albeit that interstellar radio-wave scintillation shows a
much richer range of scintillation effects than does atmospheric
optical scintillation.

Radio astronomy has a long history of exploiting propagation effects
to constrain source diameters.  Pulsars were discovered as a result of
a program of \emph{interplanetary scintillation} observations to
measure source diameters at low frequencies.  In an analogous fashion,
the SKA can exploit interstellar scintillation (ISS) to obtain source
structure information on angular scales far smaller than can be
obtained by very long baseline interferometry (VLBI) on terrestrial baselines.

\subsection{Scintillation Regimes}\label{sec:turb.regimes}

\begin{figure}
\vspace*{-2cm}
\includegraphics[scale=1.5,width=\columnwidth]{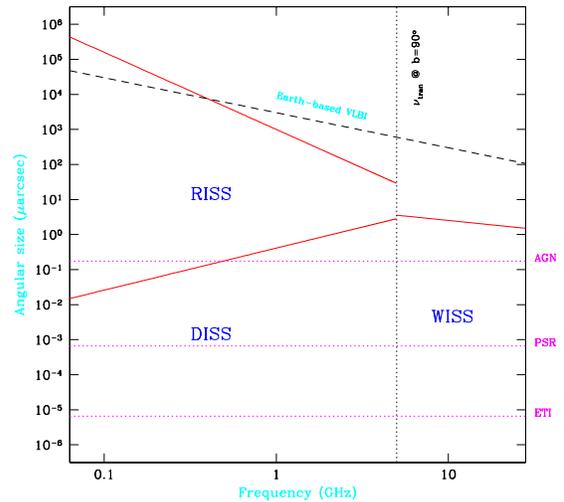}
\vspace{-1.5cm}
\caption[]{Scintillation regimes over the frequency range appropriate
to the \hbox{SKA}.  The abscissa is the frequency, and the ordinate is
a characteristic angular diameter.  Horizontal dashed lines indicate
intrinsic angular diameters appropriate for various classes of sources (\hbox{AGN},
pulsars, and ET transmitters).  Various regions are labelled as to
whether a source of a given angular diameter observed at a given
frequency could display one of the forms of \hbox{ISS}.  The
vertical dashed line shows the approximate frequency for the
transition between strong and weak ISS for observations
toward a Galactic latitude of $|b| = 90^\circ$.  The sloping line near
the top of the plot indicates the angular resolution limit for VLBI
observations over terrestrial baselines.}
\label{fig:turb.regimes}
\end{figure}

The Fresnel scale, $R_F \sim \sqrt{\lambda D}$, defines the
approximate spatial scale over which an observer receives radiation.
The character of ISS or the scintillation regime depends upon the rms
phase imposed by density fluctuations over the Fresnel scale.  If the
rms phase is comparable to or less than 1~radian, the
\emph{weak scintillation} regime is obtained; if the rms phase is
greater than 1~radian, the \emph{strong scintillation} regime is
obtained.  Figure~\ref{fig:turb.regimes} summarizes these ISS regimes
over a frequency range appropriate to the \hbox{SKA}.

\begin{description}
\item[Weak Interstellar Scintillation (WISS)]
A geometrical optics approach can be adopted, and the phase variations
result in modest focussing and defocussing of the radiation.  The
normalized rms intensity variation or \emph{scintillation index} is
less than unity.  A source will display WISS if its angular diameter
is comparable to or less than an isoplanatic angle of approximately
$R_F/D$.  WISS typically occurs for sources observed over modest path
lengths at frequencies of order 5~GHz and produce intensity
fluctuations on time scales of order hours to days and on frequency
scales that are comparable to the radio wave frequency.
\end{description}

In the strong scintillation regime, the power spectrum of the
intensity fluctuations bifurcates, and two kinds of ISS are observed.

\begin{description}
\item[Diffractive intensity scintillations (DISS)]
An rms phase of unity is obtained on the diffractive scale~$l_d$, with
$l_d \ll R_F$.  Consequently, there are many independent phase
variations within the Fresnel scale, strong constructive and
destructive interference results (Figure~\ref{fig:turb.ds1133+16}), and the (diffractive) scintillation
index saturates at unity.  In general, a wave optics approach must be
adopted.  A source will display DISS only if its angular diameter is
comparable to or less than an isoplanatic angle of approximately
$l_d/D$.  This is a stringent requirement, as $l_d \sim 10^9$~cm and
$l_d/D \sim 10^{-6}$~arcseconds for typical path lengths through the
ISM at meter wavelengths and only pulsars are known to display
\hbox{DISS}.  DISS is observed on times scales $\sim 100$~s
and frequency scales $\sim 1$~MHz, though these scales are highly
dependent on frequency, direction, and source distance and velocity.
The diffracted radiation has an angular spectrum with a characteristic
scale of order $\theta_d \sim \lambda/\ell_d$.

\begin{figure}
\includegraphics[width=\columnwidth]{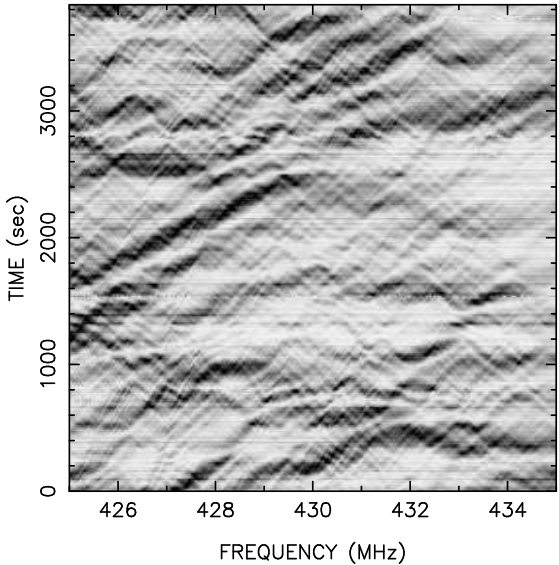}
\vspace{-1.25cm}
\caption{Dynamic spectrum for the  pulsar PSR~B1133$+$16 in the strong
scintillation regime.  The abscissa is the observing frequency, the
ordinate is time, and the gray scale shows the pulsar intensity, with
dark areas indicating strong intensity.  The time-frequency structure
is an illustration of superresolution as it includes fringes
associated with interference between radiation arriving in small
discrete bundles of ray paths. The data are from the Arecibo
Observatory (\hbox{JMC}, unpublished).}
\label{fig:turb.ds1133+16}
\end{figure}

\begin{figure}
\includegraphics[width=\columnwidth]{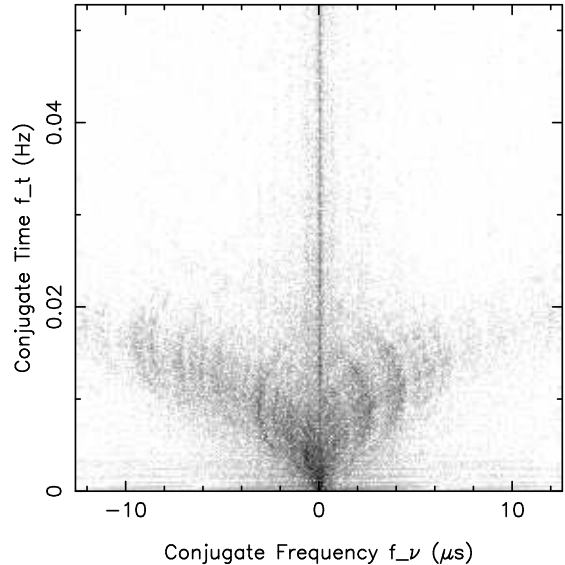}
\vspace{-1.25cm}
\caption{
The 2D power spectrum (``secondary spectrum'') of the dynamic spectrum
in Figure~\ref{fig:turb.ds1133+16}. The axes are conjugate to those in
the dynamic spectrum and thus have reciprocal units.  The gray scale
maps 5 orders of magnitude in going from black (brightest) to white
(dimmest).  The vertical feature along the $f_t$ axis at $f_{\nu} = 0$
is caused by broadband, pulse to pulse variations intrinsic to the
pulsar.  Other features include slanted and arc-like structures
(discovered in \cite{smcbgkss01}) caused by interference between
radiation contributions arriving at widely spaced angles (wide
compared to the rms or characteristic scattering angle).  These
features can be used to resolve pulsar magnetospheres and fine
structure in other sources that show \hbox{DISS}.  The resolving power
of the phenomenon is comparable to that of an interferometer whose
baseline equals the projected separation of the rays at the distance
of the effective scattering screen.  The weak arcs therefore provide
greater angular resolution than do the more prominent features in
dynamic spectra that are related to the rms scattering angle.  To use
arcs for this purpose, dynamic spectra and their secondary spectra
could be analyzed for different pulse components of a pulsar.
Alternatively, a single secondary spectrum could be synthesized that
takes an overall finite source size into account.  The resolving power is in the vicinity of~0.1--10~$\mu$s.}
\label{fig:turb.ss1133+16}
\end{figure}

\item[Refractive intensity scintillation (RISS)]  
Large scale focusing and defocusing of the radiation occurs over the
refractive scale, $\ell_r \sim D\theta_d$, implying the well known
relation, $\ell_r\ell_d \sim R_F^2$.  Geometrical optics are
sufficient, similar to \hbox{WISS}, and the (refractive) scintillation index
is less than unity.  The requirements to show RISS are more modest
than for DISS as the source diameter must be comparable to or less
than $\ell_r/D \sim \theta_d$ and $\theta_d \sim 1$~mas at~1~GHz.  In
addition to pulsars, active galactic nuclei (AGN) and masers have
exhibited \hbox{RISS}.
\end{description}

Finally, although AGN are typically too extended to display \hbox{DISS},
their angular diameters can be comparable to~$\theta_d$ meaning that
they can display angular broadening and be used to probe interstellar
density fluctuations.

\subsection{Superresolution Using DISS} \label{sec:superres}

DISS is caused by multipath scattering of radio waves from small-scale
density irregularities in the interstellar plasma. It is sensitive to
intrinsic sizes of radiation sources in much the same way that optical
scintillation from atmospheric turbulence is quenched for planets
while strong for stars.  However, interstellar scintillation differs
from the atmospheric case in that it can resolve sources at angular
resolutions much smaller than those achievable with available
apertures, including the longest baselines used in \hbox{VLBI}, even
those using space antennas.  Optical techniques such as intensity
interferometry, speckle interferometry and adaptive optics typically
only {\it restore} the telescope resolution to what it would be in the
absence of any atmospheric turbulence.

Figure~\ref{fig:turb.ss1133+16} shows the power spectrum of the
dynamic spectrum in Figure~\ref{fig:turb.ds1133+16} that reveals low-level arc like features that are caused by interference between wide-angle and weakly scattered radiation.  These features are especially promising for use in resolving radio sources that show \hbox{DISS}.

\begin{figure}
\includegraphics[width=\columnwidth]{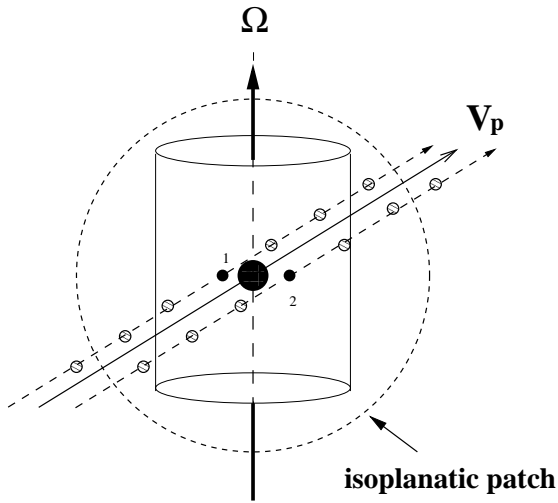}
\vspace{-1.25cm}
\caption{Depiction of pulsar geometry and the isoplanatic patch size
for \hbox{DISS}.  The cylinder indicates a portion of 
the velocity of light cylinder, the spin axis $\Omega$, and the
pulsar's transverse space velocity $V_p$.  The large filled circle on 
the spin
axis represents the neutron star while the smaller dots labelled 
1 and 2 represent two emission regions that correspond, e.g., to
two separate pulse components.  The spatial offset between the emission
regions in principle can be measured by investigating the differential
dynamic spectrum between the two pulse components.   The circles along
the dashed tracks indicate the spatial locations of the emission regions
as the pulsar moves in the plane of the sky. The size of the isoplanatic
patch is a strong function of the scattering strength, being smaller
for larger scattering.  Thus using DISS to resolve particular objects requires
flexibility in the choice of frequency.  Adequate time and frequency
resolutions are also required to sample the DISS appropriately.
}
\label{fig:turb.psr.geometry}
\end{figure}

We define the {\it superresolution regime} where
the source is unresolved by terrestrial interferometers but is sufficiently
extended to modify the \hbox{DISS}.   Let
$\ths, \thij$, and $\thiso$ be the
source size, interferometer fringe spacing, and isoplanatic DISS patch,
respectively.
By definition, two point sources separated by much less
than the isoplanatic angle  will show identical \hbox{DISS}.
The isoplanatic angle $\thiso \sim \lambda / D\thd \sim l_d/D$.
Using typical numbers
($D=1$ kpc, $\thd= 1$ mas at an observing frequency of 1 GHz),
$\thiso \sim 0.4\, \mu$arc sec.  For pulsars, whose light-cylinder
radii, $\rlc = cP/2\pi$ ($P$ = spin period) are smaller than
1 $\mu$arc sec at typical distances, we have
$\thiso \lesssim \ths \ll \thij$  and $\thiso \ll \thd$ (Figure~\ref{fig:turb.psr.geometry}).
In this case, speckle methods can achieve far better resolution
than the interferometer.  By speckle methods, we mean observations that
analyze differences in the DISS between source components, which
are sensitive to the spatial separations of those components.
Cornwell \& Narayan~\cite{cn93} and Cordes~\cite{c04a,c04b} have discussed particular superresolution
techniques in the radio context.
In optical astronomy, superresolution is not achievable
because $\thd \sim \thiso$.  However, the superresolution
regime has been identified in optical laboratory applications \cite{cmz90}.

\subsection{Intraday Variability: RISS Superresolution}\label{sec:turb.IDV}

Intraday variability (IDV) refers to the hourly to daily intensity
fluctuations exhibited by approximately 15\% of all compact, flat-spectrum
extragalactic sources at centimeter wavelengths \cite{lovelletal03}.
The phenomenon was first identified as ``flickering'' on time scales
of less than two days \cite{h84}, and variations on timescales of less
than twelve hours were reported soon after \cite{whsk86}.   It was
unclear whether the variations were intrinsic to the sources
themselves or a result of propagation effects in the \hbox{ISM}.
Initial tentative evidence in support of the former
\cite{quirrenbachetal91} was particularly concerning because it
implied source brightness temperatures in excess
of~$10^{19}$~\hbox{K}, well above the inverse Compton limit.  The
problem was exacerbated by the discovery of 40--60~min.\ variations in
PKS~0405$-$385 \cite{kedziora-chudczeretal97} for which the
variations, if intrinsic, would imply a brightness temperature
$10^{21}$~\hbox{K}.

It is now clear that the varations are not solely intrinsic to the
sources themselves but are due at least partially, if not fully, to
\hbox{ISS}.  The case for ISS hinges on two distinct observational
properties of IDV sources.  The first is the measurement of
significant time delays between widely separated telescopes in the
variability patterns of the three fastest IDV sources, PKS~0405$-$385
\cite{kedziora-chudczeretal97}, J1819$+$3845 \cite{d-tdb00}, and
PKS~1257$-$326 \cite{bignalletal03}.  The delay arises because the
intensity fluctuations caused by the scattering medium move across the
line of sight with a finite speed (which is a vector sum between the
Earth and medium's velocities), so that there is a finite delay between the intensity fluctuations observed on $\sim 5000$~km baselines. The effectiveness of this technique depends on the ability to measure significant flux density changes on timescales shorter than the time delay between the two telescopes, typically 30--600~s.   Time delay experiments have so far been limited to only the three strong sources whose variations are sufficiently strong and rapid. 

The second is that 
annual cycles in the variability timescales of the fluctuations are
observed in a number of IDV sources.  This cycle arises because the
speed of the scattering medium responsible for the ISS is comparable to Earth's orbital speed around the Sun.
The scintillation fluctuations are slowest when the Earth moves
parallel to the scattering medium and are fastest when Earth moves
antiparallel to the medium. Annual cycles have now been found for five
sources, including J1819$+$3845 \cite{d-tdb03}, B0917$+$624
\cite{rickettetal01,jm01}, and PKS~1257$-$326 \cite{bignalletal03},
and there is increasing evidence (e.g., from the MASIV survey \cite{lovelletal03}) that it is present in a number of other sources exhibiting centimeter-wavelength \hbox{IDV}. This technique has established ISS as the principal mechanism responsible for IDV for sources whose variations are too slow ($> 2$~hr) to be amenable to time delay experiments.

The variations in IDV sources generally exhibit the largest fractional
modulations at frequencies $\nu \approx 5\,$GHz.  Faster, less highly
modulated variations are observed at higher frequencies, while the
variations at lower frequencies are much slower with considerably
lower amplitude.  The change in the character of the scintillations is
interpreted in terms of a transition between weak and strong
\hbox{ISS}, with the slow variations at lower frequencies being due to
RISS and the faster variations at higher frequencies being due to
\hbox{WISS}.

The small diameters required to exhibit ISS imply uncomfortably high
brightness temperatures for IDV sources.  A source must be no more
than tens of microarcseconds in diameter to display IDV at~5~GHz for a
scattering screen at a distance of order 100~pc
(\S\ref{sec:turb.regimes}).  Brightness temperatures of order $5
\times 10^{12}$~K have been determined for B1257$-$326 and
J1819$+$3845 while the lower limit for B1519$-$273 is $5 \times
10^{13}$~K and it could be as high as $6 \times 10^{14}$~K
\cite{mk-crj00}.  While these brightness temperatures are lower than
early estimates, in which the variations were interpreted as being
solely intrinsic, they are still well above the inverse Compton limit.
Doppler beaming can reduce these estimates further, but the required
Doppler factors are as high as several hundred \cite{r94},
significantly higher than seen in existing VLBI surveys
\cite{zensusetal03,vermeulenetal03,jm03}.

Further information about source structure can be obtained by
utilizing polarization.  Even a rudimentary analysis shows that the
linearly polarized structure of J1819$+$3845 at~5~GHz consists of at
least two separate sources, separated by roughly 50~$\mu$as and
indicates their location relative to the rest of the unpolarized
emission.  Analysis of the total intensity and circular polarization
fluctuations in PKS~1519$-$273 indicates that this source consists of
a 15--35~$\mu$as core with an extremely high circular polarization of
$-3.8 \pm 0.4$\% \cite{mk-crj00}. Continuing observations of this
source indicate the intermittent generation of circular polarization
of opposite handedness in the core, suggesting that the source
undergoes small, otherwise undetectable, outbursts every few months.

The sensitivity of the SKA is such that it potentially will probe the
\emph{nanoarcsecond} scale of extragalactic sources.  It should be
capable of detecting scintillating sources as much as three orders of
magnitude fainter than those detected currently.  Assuming that these
fainter sources have incoherent synchrotron components in them, the
implied angular scales are at least one and possibly two orders of
magnitude smaller, so on the 1~$\mu$as to~100~nas scales.  One
microarcsecond corresponds to about~1~light-week at cosmological
distances, and it is on these scales that jets are expected to be
``launched.''

If there are such compact sources, the scintillation timescales would
be even faster and modulation indices potentially higher; DISS might
even be detected.
Observations of time delays between Stokes parameters will yield
differential source structure, and monitoring these delays will lead
to the evolution of these components.

\subsection{Strong Refractive Events}\label{sec:turb.ese}

A small number of pulsars have displayed ``strong fringing events'' in
their dynamic spectra, in which the normal, random appearance of the
dynamic spectrum is modified by the appearance of quasi-periodic
fringes (Figure~\ref{fig:turb.ds1133+16}).  These fringes are interpreted most naturally as the beating
of two distinct images of the pulsar.  Multiple imaging of a pulsar
can occur if there is a temporary increase in the power on refractive
scales ($\sim 10^{13}$~cm) along the line of sight, e.g., as due to a
discrete ``cloud'' of electrons drifting in front of the pulsar
\cite{cw86,cfl98}.

The fringe spacing in the dynamic spectrum can be related to the
angular separation of the multiple images formed by the strong
refraction.  The typical angular separation in the observed cases is a
few milliarcseconds \cite{cw86,wc87}.  Moreover, in order for fringes
to occur, the multiple images must be at least partially coherent.
Consequently, the strong refraction forms an interferometer with an
effective baseline of~1~\hbox{AU}, and angular resolution of~1~$\mu$as.  In
the cases observed, the fringes displayed a dependence on pulse phase,
which in turn indicated that the size of the emitting region must be
of order $10^8$~cm.  

The duty cycle of strong fringing effects, though not known with
precision, is small; it is not clear if the discrete density
structures occur along a few, specific lines of sight or if the
discrete density structures are ubiquitous but have a small filling factor.

Multiple imaging is also expected from AGN undergoing ``extreme
scattering events'' (ESE) \cite{fdjh87,rbc87,cfl98}.  Because of their
intrinsically larger diameters, DISS does not occur and dynamic
spectra are not determined.  Imaging of a source undergoing an ESE may
show the multiple imaging, if the image separation is larger than the
scattering diameter.  Unfortunately, the only VLBI observations of a
source undergoing an ESE do not included the egress of the source from
the \hbox{ESE}, when the image separation is predicted to be the largest
\cite{lazioetal00}.

\section{The Microphysics of the Interstellar
	Plasma}\label{sec:turb.physics}

For observations at~1~m wavelength over a length scale of~1~kpc, the
Fresnel scale is $R_F \sim 10^{10}\,\mathrm{cm} \sim
10^{-8}\,\mathrm{pc}$.  \emph{Scattering provides a powerful probe of
interstellar physics from parsec scales down to a a few hundred kilometers.}

Our heuristic descriptions of scattering observables and scintillation
regimes can be cast more rigorously in terms of moments of the
electric field.  Angular broadening is measured from the visibility
function, a second moment of the field, while scintillation
observables are measured from intensity correlation functions, fourth
moments of the field.  In general, analytical solutions cannot be
obtained for all desired moments of the field in all scattering
regimes, but the various moments can be shown to be related to moments
of the radio wave phase, and as such, can be related to the phase and
density fluctuation power spectra (e.g., \cite{r90}).

A useful parameterization of the spatial power spectrum of the
interstellar density fluctuations (like those in the interplanetary
medium) is a power law,
\begin{equation}
P_{\delta n_e} = C_n^2(z)\,q^{-\alpha}e^{(ql_1)^2/2}.
\label{eqn:turb.spectrum}
\end{equation}
Here $q$ is the spatial wavenumber; $l_1$ is the inner scale or
smallest length scale on which fluctuations are maintained; and
$C_n^2(z)$, which is slowly varying with distance ($z$), describes the
strength of the fluctuations along the line of sight, and is
proportional to the rms density.  Although not incorporated explicitly
into equation~(\ref{eqn:turb.spectrum}), there is presumably also an
outer scale~$l_0$ describing the largest length scales on which
density fluctuations are maintained.

\begin{figure}
\vspace*{-0.5cm}
\includegraphics[width=\columnwidth]{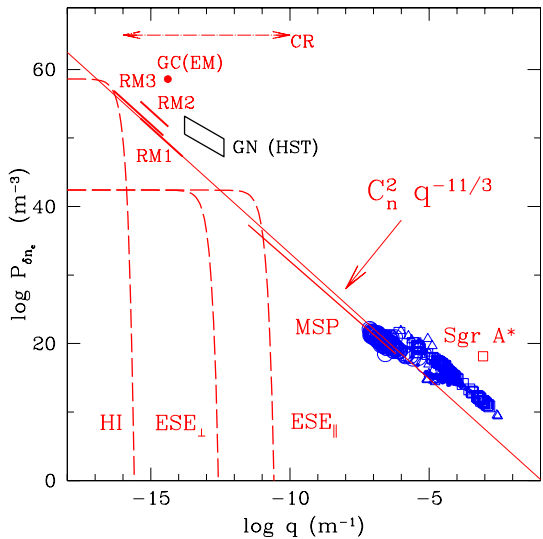}
\vspace{-1.5cm}
\caption[]{The spatial power spectrum of interstellar electron density
fluctuations as probed by various methods.  At high spatial
frequencies ($\sim 10^{-5}$~m${}^{-1}$) there are various diffractive
measurements derived from pulsar scintillation measurements and
angular broadening measurements, largely of extragalactic sources.
The box centered roughly on $10^{-10}$~m${}^{-1}$ illustrates limits
placed from millisecond pulsar (MSP) timing measurements
\cite{cwdbs90}.  The boxes labelled ``RM1,'' ``RM2,'' and ``RM3''
result from various RM studies \cite{sc86,lsc90,ms96}.  The box
labelled ``GN (HST)'' results from Hubble Space Telescope observations
of the Guitar Nebula.  The arrows at the top indicate the approximate
range of spatial scales for magnetic field fluctuations in order for
cosmic ray protons (CR) to be sufficiently scattered.}
\label{fig:turb.spectrum}
\end{figure}

A variety of studies have found that, in the solar neighborhood,
$C_n^2 \sim 10^{-3.5}$~m${}^{-20/3}$, $\alpha \approx 3.7$, $l_1
\lesssim 10^9$~cm ($\sim 10^{-9}$~pc), and $l_0 \gtrsim
10^{14}$--$10^{18}$~cm ($\sim 10^{-4}$--1~pc) \cite{ars95}.
Figure~\ref{fig:turb.spectrum} presents a unified picture of the
density fluctuations in the interstellar plasma.  The power spectrum
of the fluctuations (either density or magnetic field fluctuations) is
plotted as a function of wavenumber along with representative data.
Also indicated are the wavenumber regimes sampled by the different
techniques described in this chapter.  We now consider various aspects
and implications of this parameterization for the spatial power spectrum.

\subsection{Magnetohydrodynamic Turbulence and Interstellar Scattering}\label{sec:turb.mhd}

It is easy to show that the interstellar plasma has a large Reynolds
number.  Moreover, the density spectral index $\alpha \approx 11/3$,
close to the value expected for Kolmogorov turbulence.  This
combination, the value for $\alpha$ and the plasma Reynolds number,
suggests that the interstellar density fluctuations are the result of
a turbulent process.

There are a number of significant caveats to this conclusion.
Kolmogorov turbulence theory was developed initially for a neutral,
incompressible medium---assumptions that the ISM
clearly violates.  If interstellar scattering is an observational
manifestion of turbulence, the implied inertial range (between~$l_1$
and~$l_0$) is at least 5 orders of magnitude and potentially as much
as 10 orders of magnitude, and it is not clear how to sustain such a
large inertial range.
Moreover, strong refractive events (\S\ref{sec:turb.ese}) likely
indicate the presence of discrete density structures.

Much progress, on both the observational and theoretical fronts, has
been made in the past decade in explaining the density fluctuations
responsible for interstellar scattering in terms of turbulence.  From
the theoretical perspective, the (mean) magnetic field energy density
in the interstellar plasma is comparable to or much greater the gas
kinetic energy density or thermal pressure.  Thus, a theory for
turbulence in magnetohydrodynamics (MHD) must be used.  On the one
hand, an MHD approach seems attractive as, for instance, density
fluctuations are a natural result of Alfv\'en waves.  On the other
hand, it is probable that, at least on some spatial scales, the
nonlinear terms in the MHD equations are important and a perturbative
approach is not valid.  Over the past decade, there has been a renewed
interest in understanding MHD turbulence as it applies to interstellar
density fluctuations \cite{sg94,gs95,gs97,nb97,lg01,lg03}.  A key
aspect of a turbulent description, though, is that on the smallest
spatial scales the kinetic energy of the turbulence is dissipated and
heats the medium.  Cooling mechanisms within the medium must be
capable of radiating this heat lest a ``thermodynamic catastrophe''
result \cite{s91}.  Estimates of the dissipation heating from various
damping mechanisms can be used to constrain the nature of the
turbulence \cite{ms97}.

On the observational front, there has also been considerable progess
in establishing that there are magnetic field fluctuations
accompanying (or more likely driving) the density fluctuations
responsible for scattering.  RM and EM structure function analyses
across several degrees of sky show the structure functions to have
power-law spectra, at least on scales less than roughly 1~pc,
consistent with what is expected from a turbulent process
\cite{ms96,hgm-gdg04}.  Importantly, the RM structure function
analyses require both magnetic field and density fluctuations, as
expected if the density fluctuations result from magnetic field
fluctuations in a turbulent magnetized fluid.

\subsection{Anisotropy}\label{sec:turb.anisotropy}

In MHD turbulence, it is expected that density fluctuations would
be aligned along the field direction and potentially be highly
anisotropic.  If the density fluctuations are elongated, scattering
observables would be affected, e.g., scattering disks of sources would
be anisotropic.  Such an effect is seen in the solar wind.  

A small number of highly scattered sources do display anisotropic
scattering disks, with typical aspect ratios of a few.  The aspect
ratios are much smaller than what is expected for the density
fluctuations (for which aspect ratios of tens to hundreds may be
obtained).  However, if these aspect ratios do result from elongated
density fluctuations, the orientation of the magnetic field (and
therefore of the density fluctuations) probably varies randomly along
the line of sight with a typical correlation length comparable to the
outer scale.  If the path length through the scattering medium is
much longer than the outer scale, averaging would reduce the observed
aspect ratio well below the intrinsic aspect ratio of the density
fluctuations.

More recently, IDV observations suggest that density fluctuations in
the more general ionized interstellar medium (as opposed to
particularly heavily scattered lines of sight) may also be
anisotropic.   An axial ratio $\sim 4$:1 is derived for PKS~0405$-$385
\cite{rk-cj02} and $15:1$ for J1819$+$3845 \cite{d-tdb03}.  In the
latter case, this aspect ratio probably reflects not only scattering
but intrinsic structure within the source (i.e., jet structure).
Nonetheless, the anistropy intrinsic to the scattering medium is still
\emph{at least} 4:1 in this source.

Indeed, for two sources, Cyg~X-3 and NGC~\hbox{6334B}, the orientations of
the scattering disks change as a function of~$\lambda$
\cite{wns94,tmr98}.  Both groups identify the orientation changes as
due to the magnetic field aligning density fluctuations on different
length scales.  Notably, though, neither Sgr~A${}^*$ (e.g.,
\cite{rogersetal94,y-zcwmr94}) nor the extragalactic source
B1849$+$005 \cite{l04} show any orientation changes with wavelength,
even though the angular broadening for both has been measured over a
larger range in wavelength than for either Cyg~X-3 or
\hbox{NGC~6334B}.

\subsection{Inner Scale}\label{sec:turb.l1}

In a turbulent process the inner scale~$l_1$ is the scale on which
turbulent energy is dissipated and heats the medium.  Estimates of the
turbulent heating rate of the medium and avoidance of a thermodynamic
catastrophe rely on obtaining a robust estimate of its value.

Comparing various VLBI observations, Spangler \& Gwinn~\cite{sg90}
found evidence for an apparent change in the slope of the density
spectrum (i.e., change in~$\alpha$) on baselines $b \sim 200$~km,
consistent with detecting an inner scale of approximately this
magnitude.  They attribute the inner scale to either the ion inertial
length or Larmor radius.

\subsection{Cosmic Ray Propagation and MHD Turbulence}\label{sec:turb.g-ray}

Galactic cosmic rays are observed to have an isotropic arrival
direction and have an energy spectrum that is a continuous power law
over the approximate energy range of~1 to~$10^6$~GeV per nucleon.
(The power law may extend to lower energies, but solar modulation
becomes increasingly important at lower energies.)
Jokipii~\cite{j88,j99} has argued that these characteristics of cosmic
rays require a wide-band spectrum of magnetic field fluctuations.

The absence of any ``breaks'' or ``bumps'' in the spectrum as well as
an isotropic arrival direction indicate that a typical cosmic ray
undergoes many scatterings during its lifetime.  Cosmic rays are
scattered most effectively by magnetic field fluctuations on length
scales near their gyroradii.  In this energy range, the relevant
gyroradii are $10^{12}$--$10^{18}$~cm, suggesting that the spectrum of
magnetic field fluctuations encompasses at least this range of length
scales.  As described above, there is modest evidence from RM
structure function analyses for magnetic field fluctuations at the
upper end of these length scales, while refractive effects and DM
monitoring programs can probe density fluctuations at the lower end of
these length scales.

The large number of scatterings that cosmic rays experience also makes
them of limited utility in studying the magnetic field fluctuations,
other than indicating that magnetic field fluctuations may exist.  The
diffuse Galactic gamma-ray emission at energies $E > 100$~MeV results
from the interaction of cosmic rays with the interstellar matter and
radiation fields \cite{b89,fichteletal89} and suffers essentially no
absorption or scattering over Galactic distances.  As such it is a
powerful diagnostic of the Galactic distribution and propagation of
cosmic rays.  

Typical models for the diffuse $\gamma$-ray emission utilize
expressions like equation~(\ref{eqn:turb.g-rayI}) in which $n_n$ is
approximated by the hydrogen distribution~$n_H$ and $q_\gamma$ is
taken to be constant or slowly varying.  These models have proven
quite successful in reproducing the large-scale features---an
enhancement in the inner Galaxy and spiral-arm tangents---seen in maps
of $I_\gamma$ (e.g., \cite{strongetal88,bertschetal93,hunteretal97}).
In such models, variations in $I_\gamma$ arise essentially from
variations in $n_H$.  However, there are departures of the data from
the model. In some famous cases, these are due to discrete sources
(Geminga, Crab pulsar, Vela pulsar) while in others, they are
essentially unidentified ``hot spots'' in the gamma-rays.

``Hot spots'' and other variations in $I_\gamma$ might also result
from fluctuations in $q_\gamma$.  In particular, magnetic fluctuations
can act not only to scatter cosmic rays, but also to confine them
\cite{m77}.  Therefore, enhancements in $I_\gamma$ could result
either from an increase in the ambient matter density or from a longer
cosmic-ray residence time in regions of enhanced magnetic turbulence
or both.  If the density fluctuations responsible for interstellar
scattering result from the same magnetic fluctuations responsible for
cosmic-ray confinement, as appears likely on at least the largest
scales, radio-wave scattering data can be used to \emph{predict}
$q_\gamma$.  Conversely, \emph{gamma-ray observations could provide an
additional indirect probe of magnetic field fluctuations on length
scales smaller than those probed by Faraday rotation variations and
complementary to those probed by radio-wave scattering observations
and the energy spectrum of Galactic cosmic rays} (Figure~\ref{fig:turb.spectrum}).

To date, little has been done to test these ideas, due at least in
part to the fairly coarse angular resolution of $\gamma$-ray
telescopes.  In comparison to the sub-arcsecond resolution of the
radio telescopes typically used to study interstellar scattering, past
$\gamma$-ray telescopes have had angular resolutions of a few degrees
or worse.  More problematic, these angular resolutions also have been
fairly coarse when compared with the diameter of a typical scattering
region over Galactic distances.  For instance, a region 1~pc in
diameter at a distance of~8~kpc (comparable to the mean free path
between intense scattering regions \cite{cwfsr91}) subtends an angle
of only $0.5^\prime$.  The effects we describe may nonetheless be
detectable with future $\gamma$-ray instruments with higher angular
resolutions operating during the SKA's lifetime.

Not only could comparisons of radio and $\gamma$-ray observations
yield information about the magnetic field on sub-parsec scales, but
radio observations could be used to improve models for the Galactic
distribution of the diffuse $\gamma$-ray emission.  

The diffuse Galactic $\gamma$-ray emission can (probably does!)
obscure faint point and small-diameter sources (neutron stars, black
holes, $\gamma$-ray blazars, young supernova remnants, or even an
as-yet unidentified class of $\gamma$-ray emitters).  For example,
there are presumably more Geminga-like $\gamma$-ray pulsars in the
Galaxy.  Separating more distant and/or intrinsically weaker
$\gamma$-ray point sources from the diffuse emission requires better
understanding of the diffuse emission, particularly on smaller angular
scales.  Indeed, the detection of Geminga itself may have been
somewhat fortuitous in that it was both nearby and toward the galactic
anticenter, where the diffuse emission is less intense.

\section{Distance Determinations from Interstellar Scattering}\label{sec:turb.distribute}

The conventional factorization of the density spectrum allows for
different lines of sight to have similar microphysics (wavenumber
scaling) but differing scattering strengths (coefficient~$C_n^2$).
Values for \SM, the line-of-sight integral of~$C_n^2$, can be estimated
from angular broadening measurements of extragalactic sources or
Galactic sources (typically pulsars or masers), pulse broadening
measurements for pulsars, or scintillation bandwidth measurements for
pulsars.  In general, only angular broadening measurements of
extragalactic sources yield SM directly.  For the other kinds of
measurements, a line-of-sight weighting of the scattering material
must be taken into account in order to extract \hbox{SM}; for angular
broadening of Galactic sources, scattering material close to the
observer is most strongly weighted while for pulse broadening and
scintillation bandwidth scattering material mid-way between the pulsar
and observer is most strongly weighted.

Comparisons of SM along different lines of sight yield the angular
distribution of density fluctuations, which can be inverted to obtain
a spatial distribution, while comparisons of SM on similar lines of
sight (e.g., pulsar vs.\ AGN) constrain the \emph{radial} distribution
of density fluctuations.  For pulsars, in principle, one also can
compare the magnitude of angular broadening to the pulse broadening or
scintillation bandwidth to obtain the line-of-sight weighting
directly.

\subsection{The Galactic Distribution of the Interstellar
	Plasma}\label{sec:turb.galactic}

\begin{figure}
\includegraphics[width=\columnwidth]{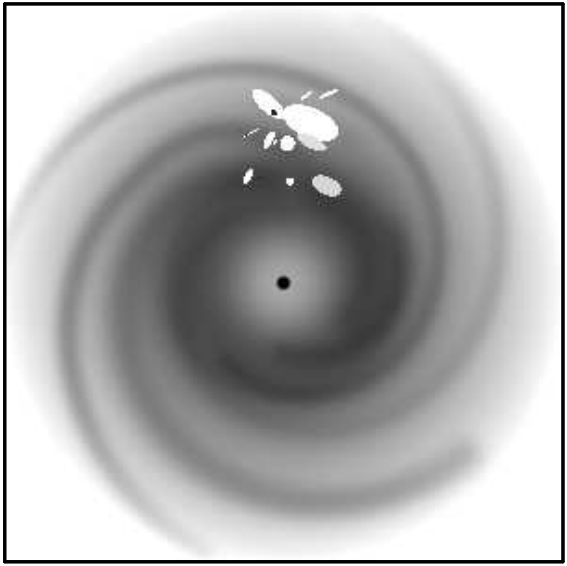}
\vspace{-1.25cm}
\caption[]{The Galactic distribution of the free electron density, as
described by the NE2001 model.  Shown is a face-on view of the Galaxy
with the Galactic center at the center.  Three large-scale components
of the model---the inner Galaxy component centered on the molecular
ring, the spiral arms, and the diffuse component---are all clearly
visible.  Also visible are a number of ``voids,'' such as the Local
Bubble, some of the mesoscale structures required in the model in
order to reproduce various pulsar or scattering observables.  The Sun
is near the black dot in the center of the Local Bubble in the upper
portion of the figure.}
\label{fig:turb.ne2001}
\end{figure}

Although measurements of only SM can be used to obtain the angular
distribution of density fluctuations, they are much more powerful when
combined with measurements of pulsar DMs, and pulsar parallaxes to
produce three-dimensional models for the Galactic distribution of the
interstellar plasma.  The most recent effort is the NE2001 model (so
named because it is based on data obtained up to~2001,
\cite{cl02,cl03}, Figure~\ref{fig:turb.ne2001}), which uses roughly 1200 pulsar DMs, 300
measurements of \SM, and~100 measurements of DM-independent pulsar
distances (e.g., parallaxes or pulsar-supernova remnant
associations).  Not only does this model describe the Galactic
distribution of the interstellar plasma, it can be used to predict the
distances of pulsars from their DMs.

The NE2001 model (like many of its predecessors) describes the Galaxy in
terms of large-scale components---a thin disk, a thick disk, and
spiral arms.  Although less true than in some previous models, the SM
and DM measurements still play a largely complementary role in the
NE2001 model.  Most pulsars are relatively faint so their DMs probe
the interstellar plasma only relatively nearby ($\sim 1$~kpc); most
measurements of SM are toward extragalactic sources so the lines of
sight are relatively long ($\sim 10$~kpc).  The combination of the DM
and SM values thus allows for the possibility of a global model.  With
the increasing number of distant pulsars and their DMs, particularly
from the Parkes multi-beam survey, the NE2001 model is perhaps more
equitable in the balance between DM and SM for long lines of sight,
though the SM values remain important.

A novel feature of the NE2001 model is the systematic introduction of
``clumps'' and ``voids,'' regions of enhanced or decreased electron
density and/or scattering.  Previous models had contained a localized
enhancement (clump) due to the Gum Nebula because it is so close to
the Sun, but the current number of measurements of DM and SM is
becoming sufficient that not only large-scale features (e.g., diffuse
ionized disk, spiral arms) can be modeled but also mesoscale
structures (e.g., H\,\textsc{ii} regions).  On a limited number of
lines of sight, reasonable agreement between a modeled and observed
quantity (DM and/or SM) could be obtained only by inserting either a
clump or a void.

The SKA will provide qualitatively new means of constructing global
models.  One of the Key Science Projects for the SKA is a pulsar
census of the Galaxy (Kramer et al., this volume; Cordes et al., this
volume), which will necessarily include the DMs for all of the pulsars
found.  It is expected that such a census will increase the number of
pulsars by a substantial factor ($\sim 5$).  We expect that the number
of pulsar parallaxes will also increase by a similar amount.  The
number of pulsars will be sufficiently large that there should be
several per degree along the Galactic plane.  Consequently, the lines
of sight to distant pulsars will have a reasonable probability of
intersecting an H\,\textsc{ii} region even over Galactic distances
($\sim 10$~kpc).

The density of pulsars on the sky will decrease with increasing
Galactic latitude of course, but these can be supplemented with
scattering measurements and IDV observations. Assuming that the
density of IDV sources on the sky is sufficiently high, the
characteristic frequency at which IDV is strongest can be determined
as a function of Galactic latitude, in a manner analogous to
observations of interplanetary scintillation at lower frequencies
(e.g., \cite{rd-s75}).

We suggest that the number of measurements provided by the SKA will be
sufficiently large that it will be possible to trace the Galaxy's
structure, or at least its spiral structure, in a self-consistent
manner, rather than by imposing it as has been done for the NE2001 and
previous models.  The methodology would be to insert clumps of
increased electron density, representing H\,\textsc{ii} regions, along
the line of sight to pulsars and extragalactic sources.  The clumps
would be inserted in a parsimonious manner so as to minimize the
difference between observed and predicted DMs and SMs with the minimum
number of clumps.  As spiral arms are not smooth structures, voids may
also need to be inserted in a similar manner.  Moreover, combining
scattering measurements, e.g., pulsar pulse broadening and angular
broadening, would constrain the \emph{radial} distribution of
scattering (e.g., \cite{cgdb02}).

A second Key Science Project (see chapter by Gaensler, this volume),
is the production of a grid of extragalactic sources with measured
RMs.  Although not used to date in constructing models for the
Galactic distribution of the interstellar plasma, such a grid offers
exciting possibilities for expanding the NE2001 model (or its
successors) to describe the magnetoionic medium in terms of both $n_e$
and~$B$.

\subsection{Local Interstellar Medium}\label{sec:turb.lism}

IDV sources are excellent probes of the scattering material within
roughly 10--50~pc of the Sun.  The prominence of such local scattering
material in IDV scattering would, at first sight, appear to be
surprising, since pulsar radiation tends to be scattered at greater
distances. However, for WISS and \hbox{RISS}, the finite angular
diameters of IDV sources produce a weighting with nearby scattering
material being most important.  This effect arises because WISS and
RISS tend to be dominated by scattering material whose distances are
such that the angular diameter of the first Fresnel zone closely
matches the source angular diameter \cite{cf87,cfrc87}.  The
combination of a large number of IDV measurements along with pulsar
scattering measurements and parallaxes should yield similar
qualitative advances in the description of the local \hbox{ISM}.

\section{Intergalactic Scattering}\label{sec:turb.igm}

Radio-wave scattering is a generic process that occurs whenever radio
waves pass through a medium containing density fluctuations.  To date,
observable effects have been seen from the Earth's ionosphere (e.g.,
ionospheric scintillations), the interplanetary medium (e.g.,
interplanetary scintillations and spectral broadening), and the
interstellar medium (see above).  That a small number AGN display
intraday variability and a small number of gamma-ray burst afterglows
display interstellar scintillation indicates that at least some lines
of sight through IGM are not scattered appreciably.  If these lines of
sight were scattered, these sources would be too heavily broadened to
display interstellar scintillation.

Nonetheless, the IGM would appear to have the characteristics
necessary for intergalactic scattering to be obtained.  For redshifts
$z \lesssim 6$, the absence of a Gunn-Peterson trough in quasar
spectra indicates that the IGM is largely ionized, and recent
far-ultraviolet and soft X-ray observations of highly ionized species
(e.g., O~\textsc{vi}) along the line of sight to various low-redshift
quasars may be a direct detection of the ionized IGM
\cite{oegerleetal00,tsj00,rswsk01,shsso01,sstr02}.  In the prevailing
``cosmic web'' scenario for large-scale structure formation, most of
the baryons are in filamentary structures which are themselves
permeated by shocks.  Thus, it is expected that the IGM is both
ionized and ``turbulent.''  Moreover the long path lengths
through the IGM ($> 100$~Mpc) as compared to the ISM ($\sim
1$--10~kpc) may compensate for the lower density ($\sim
10^{-7}$~cm${}^{-3}$ vs.\ 0.025~cm${}^{-3}$, respectively).

Detection of intergalactic scattering would be a powerful probe of the
\hbox{IGM}, as it would be caused by the majority of the baryons.
Observations of the Ly$\alpha$ forest probe mostly the neutral
component, which represents only about $10^{-5}$ of the mass.  The
metallicity of the IGM is highly uncertain, so the ionized gas traced
by the far-ultraviolet and soft X-ray observations represents a small
and uncertain fraction of the total mass.

Cordes \& Lazio~\cite{cl04} have considered intergalactic scattering
from the general \hbox{IGM}, expanding on previous treatments of
scattering from intra-cluster media \cite{hs79}.  In analogy to
interstellar scattering, dispersion and scattering measures from the
IGM can be defined and both diffractive and refractive scattering
effects can be described.  We summarize their conclusions briefly.

The expected DM through the IGM is $\mathrm{DM} \gtrsim
1000$~pc~cm${}^{-3}$ for sources with redshifts $z \gtrsim 4$.  A
similar estimate has been found by Palmer~\cite{p93}.  This value is
comparable to path lengths of order a few kiloparsecs through the
\hbox{ISM}.

The expected SM through the IGM depends upon location of the bulk of
the scattering material.  If the bulk of the scattering arises in
galaxies like the Milky Way (rather than the diffuse IGM),
$\mathrm{SM} \sim 10^{-4.5}$~kpc~m${}^{-20/3}$.  Estimating the
scattering measure in the diffuse IGM requires assumptions about the
distribution and power spectrum of density fluctuations in the
\hbox{IGM}.  Assuming that the density fluctuations are in
``cloudlets,'' (as in the ISM), they estimate that scattering measures
as large as $\mathrm{SM} \sim 10^{-3}$~kpc~m${}^{-20/3}$ may be
obtained.  In the latter case, scatter broadening diameters of order
1~mas are implied.  As in the interstellar case, there is a
geometrical weighting factor for the distribution of density
fluctuations along the line of sight, with material closest to the
observer having the largest weight.  For intergalactic scattering,
there is an additional wavelength-dependent weighting, that also
favors scattering material close to the observer; at higher redshifts,
the wavelength of the propagating radiation is shorter and scattering
is generally less effective.

\begin{figure}
\includegraphics[angle=-90,width=\columnwidth]{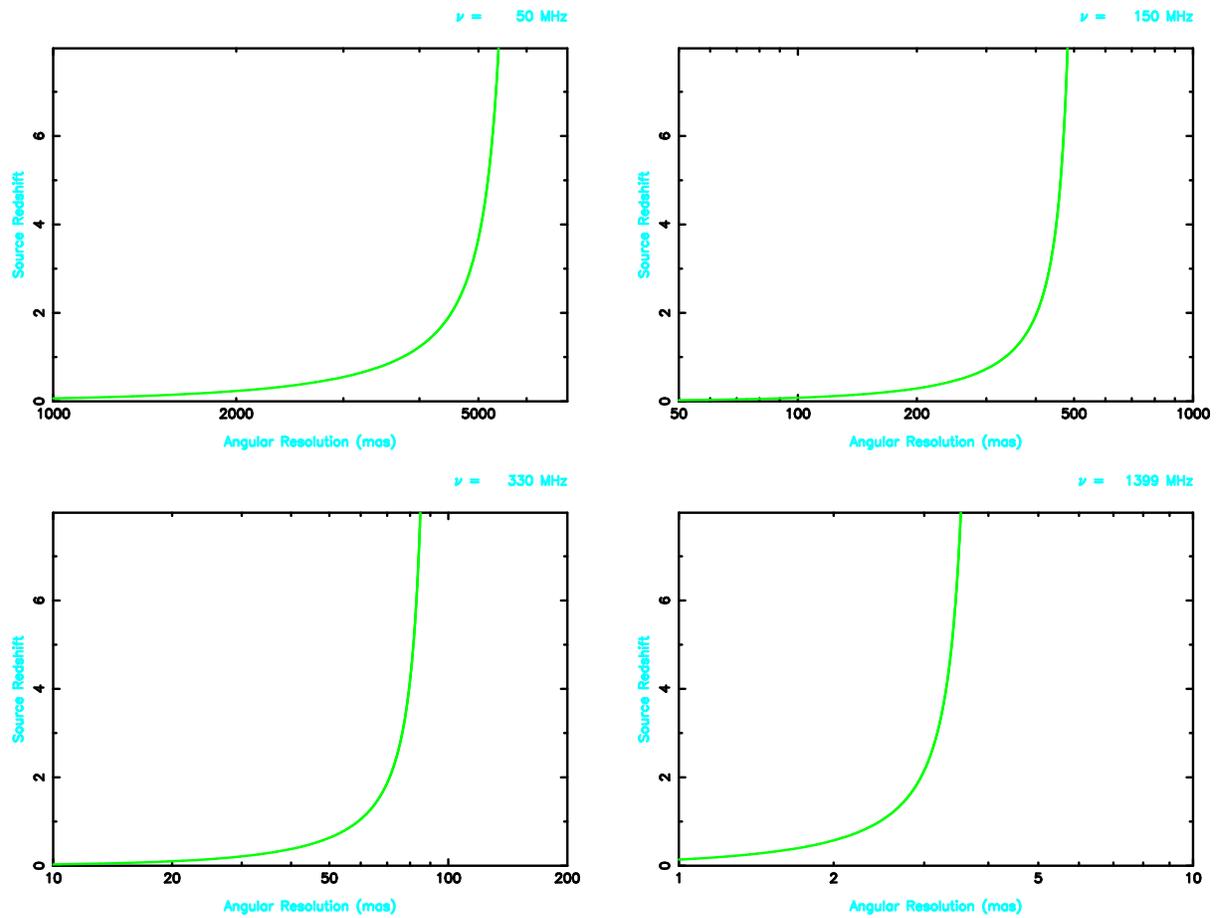}
\vspace{-1cm}
\caption[]{Source redshift required to produce angular broadening of a
given magnitude at four typical observation frequencies.  Clockwise
from upper left, the four frequencies are 0.05, 0.15, 0.33,
and~1.4~GHz.  The curves shown assume that the scattering medium is
intergalactic clouds with $\Omega_{\mathrm{H\,II}} \sim 0.04$.}
\label{fig:turb.igm}
\end{figure}

Figure~\ref{fig:turb.igm} shows predictions for the amount of angular
broadening expected at selected frequencies, based on a model in which
the bulk of the IGM is the form of electron density ``cloudlets.''
At~1.4~GHz an angular resolution better than 4~mas is required to
detect intergalactic scattering.  Such an angular resolution is at the
limit of what can be achieved with terrestrial baselines.  Reliable
detection of intergalactic scattering at~1.4~GHz probably will
require space-based \hbox{VLBI}, though the SKA will prove valuable in
increasing the senstivity of such future VLBI arrays.

At~0.33~GHz an angular resolution better than 80~mas is required.
This requirement is easily within the capabilities of existing VLBI
networks.  For instance the highest angular resolution of the VLBA at
this frequency is 25~mas.  The key difficulty with existing VLBI
arrays is their relative lack of sensitivity at this frequency.  Even
if only a small fraction (5--10\%) of the SKA is distributed on
continental/intercontinental baselines, its sensitivity could exceed
existing VLBI arrays at this frequency by as much as a factor
of~$10^3$.  Such a large increase in sensitivity would produce a
significant increase in the number of sources that could be searched
for scattering from the IGM and enable robust statistical analyses,
e.g., for trends with redshift.

At~0.15~GHz the required angular resolution to detect intergalactic
scattering is 500~mas, equivalent to baselines of order 1000~km.  It
is not yet clear if the SKA will operate at~0.05~GHz, but, if it should,
an angular resolution better than 5\arcsec, equivalent to baselines
of~250~km, is required.  It is not clear that sufficiently compact
objects exist at cosmological distances to be detected at~0.05~GHz,
though, at a redshift of~5, for example, the rest frame emission would
occur at~0.3~GHz.  At both frequencies, the required baselines are
well within the nominal goals for the \hbox{SKA}.

\emph{Thus, detection of possible intergalactic scattering requires
the SKA to have intercontinental baselines at~0.33~GHz or baselines of
order 1000~km below~0.2~GHz or both.}

\onecolumn
\twocolumn

\section{The Variable (Scintillating) Sky and SKA Synthesis
Observations}\label{sec:turb.problems}

At the sensitivity levels reached by the \hbox{SKA}, we may expect to
see many scintillating radio sources within the field of view.  It is
been often stated, and generally believed, that variability will
severely limit the dynamic range of synthesis images. We consider two
extreme cases to illustrate why this is not necessarily so.

The most frequently encountered case will be that the total average
flux density of all scintillating sources in the field of view during
the observation is small fraction of the total flux density
contributed by all sources.  Consequently, the scintillating sources
cannot have a large effect on the self-calibration solution.

In the other extreme case, the average flux density of the
scintillating source dominates the total flux density in the field of
view.  After the light curve of the scintillating source has been
determined, it can be subtracted from the visibility data in a
time-dependent manner.  Standard imaging and self-calibration can be
applied to the residual visibility data.

\section{Why the SKA and What is Needed?}\label{sec:turb.summary}

An important aspect of many of kinds of observations relevant to
superresolution imaging and turbulent phenomena is that they represent
a class of \emph{non-imaging} analyses that the SKA should be capable
of performing.  In general, these observations require that some
(large) fraction of the SKA's collecting area be capable of being
operated in a phased-array mode, in which the voltages from the
individual collectors are delayed appropriately and then summed.  

Exploiting the SKA to study the structure of both extragalactic
sources and pulsars as well as of the magnetoionic Galactic
interstellar medium requires that a variety of observables be
measurable.  In many cases these observables are similar to what is
needed to exploit observations of transient radio sources (Cordes et
al., this volume) and GRB afterglows (Weiler, this volume), and, as
pulsars form an important target, similar to the requirements for
exploiting observations of pulsars (Cordes et al., this volume; Kramer
et al., this volume).

\begin{description}
\item[Dynamic spectra]
Probing source structure with superresolution as well as the
scattering properties of the ISM requires multiepoch dynamic spectra
(Figure~\ref{fig:turb.ds1133+16}).  In general for probing source
structure, the key promise of the SKA is its sensitivity.  Many of the
relevant observations can be conducted at the present time or in the
near future, but only on a small number of sources, so it is not clear
to what extent these sources represent typical or representative
examples.  The utility of dynamic spectra for a source is that, if
there is sufficient sensitivity to form dynamic spectra as a function
of time, modifications in the DISS characteristics in the differential
dynamic spectra indicate the presence of structure on sub-$\mu$as
angular scales.  For some applications, the dynamic spectra should be
obtained for all Stokes parameters.

For pulsars, dynamic spectra formed a function of pulse phase can be
used to assess the presence of multiple emitting regions within the
pulsar magnetosphere (Figure~\ref{fig:turb.psr.geometry}).  For GRB
afterglows, a similar analysis could be used to detect and monitor the
formation and evolution of components within the emitting shock
regions.  For \hbox{AGN}, if they do display \hbox{DISS}, the relevant
scales probed are comparable to those on which jet formation is
thought to occur near the central engine.  For the case of \hbox{AGN},
dynamic spectra observations will complement those of future missions
such as Constellation-X and \hbox{GLAST}, which will observe the
highly-energetic emission from the bases of AGN jets.

Particularly for the case of pulsars, dynamic spectra also can be
utilized to extract parameters such as the scintillation bandwidth,
scintillation time scale, and scintillation velocity, and the arc
phenomenon (Figure~\ref{fig:turb.ss1133+16}) seen in pulsar dynamic
spectra can be used to constrain the electron density power spectrum.
Although these parameters can be extracted from the current pulsar
census with existing telescopes, the much larger pulsar census that
the SKA will perform opens up the possibility of qualitatively new
analyses in determining both the Galactic (\S\ref{sec:turb.galactic})
and line-of-sight (\S\ref{sec:turb.distribute}) distributions of
scattering material.

These observations of dynamic spectra require time resolutions $\sim 1$~s
and frequency resolutions $\sim 1$--100~kHz over wide bandwidths.  In
order to obtain dynamic spectra as a function of pulsar pulse phase,
sampling times of~10 to~100~$\mu$s are needed.

\item[Pulse broadening]
Measurement of pulse broadening of pulsar pulses enables
characterization of SM along heavily scattered lines of sight,
complementing those for which dynamic spectra can be used, and
constrains the wavenumber spectrum for $\delta n_e$ and the
distribution along the line of sight of scattering material.

As with dynamic spectra, pulse broadening can be determined for a
small number of pulsars with existing telescopes, but the SKA offers
the promise of a greatly increased number of pulsars toward which
pulse broadening can be determined.  In particular, measurements of
pulse broadening in the the much larger pulsar census provided by the
SKA are important for the possibility of qualitatively new analyses in
determining the Galactic distribution of the interstellar plasma
(\S\ref{sec:turb.galactic}).  ``Clumps'' in the interstellar plasma,
as introduced in the NE2001 model, produce enhanced pulse broadening
and so may enable the Galactic spiral arms to be traced, in part, by
pulse broadening.

These observations require time resolutions, channelization, and post
processing identical to that needed for pulsar timing
(Cordes et al., this volume).

\item[Angular broadening]
Probing the Galactic (and intergalactic?) and line-of-sight
distributions of scattering material as well as comparing radio-wave
and $\gamma$-ray observations requires high angular resolution in
order to measure angular broadening.

One of the strengths of angular broadening measurements toward AGN is
that it provides a direct measure of the scattering strength along the
line of sight.  Other measures of scattering, e.g., pulse broadening,
contain line-of-sight weightings for the scattering material.  As for
pulse broadening, angular broadening is a powerful probe of the
Galactic plasma and, in terms of a qualitative improvement for the
Galactic interstellar plasma (\S\ref{sec:turb.galactic}), a large
census of \SM\ values for \hbox{AGN} would complement the large pulsar
census that the SKA will perform.  Moreover, pulsars will not be
detectable over cosmological distances, while AGN clearly are, so
angular broadening offers the promise of a direct probe of the
IGM (\S\ref{sec:turb.igm}).

For angular broadening measurements, the most important aspect is
transcontinental or intercontinental baselines.  As described in
\S\ref{sec:turb.igm}, intercontinental baselines at~0.33~GHz or
baselines of order 1000~km below~0.2~GHz or both are minimal
requirements for angular broadening studies.

\item[Light curves]
In many current concepts, the SKA is envisioned as having a ``core''
and ``outlying stations.''  Not all observations will be able to make
use of the entire SKA, e.g., use of the outlying stations produces too
low of a surface brightness sensitivity for some kinds of
H\,\textsc{i} observations.  Thus, a small number of outlying stations
could be used to form a sub-array and tasked to measure the flux
density of a variety of objects, such as to monitor a large number of
AGN for IDV or search for ESEs.
\end{description}

\begin{table*}
\caption{Requirements for the SKA\label{tab:turb.require}}
\begin{tabular}{lcc}
\noalign{\hrule\hrule}
Item     & Requirement   & Motivation \\
\noalign{\hrule}
Time Resolution      & 10--100~$\mu$s & dynamic spectra \\
                     &                & pulse broadening \\
Frequency Resolution & 1--100~kHz     & dynamic spectra \\
Configuration        & phased-array   & pulsar observations \\
                     & operations possible & \\
Configuration        & sub-arrays possible 
                                      & light curves\\
Configuration        & $> 1000$~km baselines
                                      & angular broadening \\
\noalign{\hrule}
\end{tabular}
\end{table*}

\end{document}